\documentclass[11pt,a4paper]{article}
    \usepackage{jheppub}

\usepackage{bbold}

\newcommand{\pt}[1]{\left(#1\right)}
\newcommand{\pq}[1]{\left[#1\right]}
\newcommand{\pg}[1]{\left\{#1\right\}}

\newcommand{\num}{\addtocounter{equation}{1}\tag{\theequation}}
\newcommand{\pmat}{\begin{pmatrix}}
\newcommand{\fpmat}{\end{pmatrix}}
\newcommand{\eq}{\begin{equation}}
\newcommand{\feq}{\end{equation}}
\newcommand{\cas}{\begin{cases}}
\newcommand{\fcas}{\end{cases}}

\newcommand{\mbf}[1]{\mathbf{#1}}

\newcommand{\eqarray}{\begin{eqnarray}}
\newcommand{\feqarray}{\end{eqnarray}}

\newcommand{\diag}[1]{\operatorname{diag}\pt{#1}}

\newcommand{\lagr}{\mathcal{L}}

\newcommand{\Imm}{\,\text{Im}\,}

\newcommand{\al}{\alpha}

\newcommand{\be}{\beta}

\newcommand{\ve}{\varepsilon}

\newcommand{\La}{\Lambda}
\newcommand{\m}{\mu}
\newcommand{\n}{\nu}
\newcommand{\om}{\omega}
\newcommand{\Om}{\Omega}

\newcommand{\si}{\sigma}

\newcommand{\Si}{\Sigma}

\newcommand{\half}{\frac{1}{2}}

\newcommand{\quar}{\frac{1}{4}}

\newcommand{\ft}[2]{{\textstyle\frac{#1}{#2}}}
\newcommand{\nn}{\nonumber}
\def\ii{{\rm i}}

\def\be{\begin{equation}}
\def\ee{\end{equation}}
\def\bea{\begin{eqnarray}}
\def\eea{\end{eqnarray}}

\title{ More on microstate geometries of 4d black holes }
\author{M. Bianchi$^{1}$,  J.F. Morales$^{2}$, L. Pieri$^{1,3}$ and N. Zinnato$^{1}$}
\affiliation{
   $^1$ Universit\`a di Roma Tor Vergata and I.N.F.N, Dipartimento di Fisica, 
\\ Via della Ricerca Scientifica, I-00133 Rome, Italy\\ 
$^2$ I.N.F.N - sezione di Roma 2\\
and Universit\`a di Roma Tor Vergata, Dipartimento di Fisica\\
Via della Ricerca Scientifica, I-00133 Roma, Italy\\ 
$^3$ center for Research in String Theory,  School of Physics and Astronomy, \\
Queen Mary University of London, \\Mile End Road,  London, E1 4NS, United Kingdom \\}
\emailAdd{bianchi@roma2.infn.it,morales@roma2.infn.it}
\emailAdd{lorenzo.pieri@roma2.infn.it,natale.zinnato@gmail.com}
\abstract{ We  construct explicit examples of microstate geometries of four-dimensional black holes that lift to smooth horizon-free geometries in five dimensions. Solutions  consist of half-BPS D-brane atoms distributed in $\mathbb{R}^3$. Charges  and positions  of the D-brane centers are constrained  by the bubble equations and  boundary conditions  ensuring  the regularity of the metric and the match with the black hole geometry.  In the case of three centers, we find that the moduli spaces of solutions includes disjoint one-dimensional components of (generically) finite volume.  
  }


\preprint{PREPRINT}

\begin{document}
\maketitle

\pagebreak
 
\section{Introduction}

Black holes are  classical solutions of  Einstein's equations with  curvature singularities  hidden behind event horizons. According to the ``no-hair theorem'', the  solutions are unique in four dimensions, once the mass, charge and angular momentum are specified. At the classical level black holes are absolutely black and have zero statistical entropy  $S = \log(1) = 0$. 
In a quantum theory however a black hole radiates  as a black body with a finite temperature and an entropy  given by one quarter of the area of its event horizon.  To  explain the microscopic origin of this entropy remains a primary task for any serious contender to a quantum theory of gravity.  
    
  In string theory, black holes can be realised in terms of D-branes intersecting in the internal space.  
  The micro-states can be represented (and counted) in terms of excitations of  the open strings connecting the building brane bits. Alternatively, one may think of the geometry generated by the excited brane state as the gravity representation of the micro-state.  Since the micro-state geometry describes a pure state with zero entropy,  it should have no horizon. This line of ideas motivates the ``fuzzball'' proposal that associates to every black hole micro-state a regular and horizon-free solution of classical gravity. The solutions, known as ``fuzzballs'' or ``micro-state geometries'',  share with the would-be black hole the mass, charges and angular momentum but differ from it in the interior \cite{Mathur:2005zp, Skenderis:2008qn, Mathur:2008nj}.   The black hole horizon and its entropy arise from a coarse graining superposition of the micro-state geometries.  

    In the last years, a large class of four and five dimensional black hole micro-state geometries  have been produced \cite{Bena:2005va,Bena:2007kg,Berglund:2005vb,Bena:2006kb,Bena:2007qc,Lunin:2012gp,Giusto:2013bda,Bena:2015bea,Bena:2016agb,Bena:2016ypk}. The micro{-}state geometries are typically coded in
     smooth, horizon-less geometries with no closed time-like curves (CTC's) in five or six dimensions. This is the best one can achieve, since no-go theorems in four dimensions  exclude the existence of non-singular asymptotically flat soliton solutions\footnote{Regular solutions  with AdS asymptotics have been recently found in \cite{Lunin:2015hma,Pieri:2016cqz}}.   This is not the case in five or higher dimensions where the existence of Chern-Simons interactions and  spatial sections with non-trivial topologies circumvent the no{-}go result \cite{Gibbons:2013tqa}. From a four-dimensional perspective, the finiteness of the higher dimensional Riemann tensor (and its derivatives) results into a finite effective action with  curvature divergences  compensated by  the singular behaviour of the scalars and gauge fields.   

   A black hole with finite area in four dimensions  can be realised in several different frames.  Popular choices include bound states of  D1-D5-KK-p, of D0-D2-D4-D6 branes \cite{Denef:2000nb,Denef:2002ru,Bates:2003vx,Dall'Agata:2010dy,Shih:2005qf,Bena:2005ni,Gaiotto:2005xt}  or of intersecting D3-branes, wrapping  three cycles in $T^2\times T^2\times T^2$  \cite{Bertolini:2000ei,Bianchi:2016bgx,Pieri:2016pdt}.  After reduction down to four dimensions, the solution  can be viewed as a supersymmetric vacuum of the ${\cal N}=2$  truncation of  ${\cal N}=8$ supergravity involving the gravity multiplet and three vector multiplets  characterizing the complex structures of  the internal torus. In a microscopic world-sheet description of the  D3-brane system  \cite{Bianchi:2016bgx,Pieri:2016pdt}, the harmonic functions characterising the gravity solution are sourced by disk diagrams with a boundary ending on a single, two or four different branes  \cite{Bianchi:2016bgx,Giusto:2009qq,Giusto:2011fy}.  Higher multipole modes are generated by extra insertions of untwisted open string fields on the disk boundaries.  The micro-state multiplicities  can be computed by counting open strings in the D1-D5-p-KK system or vacua  in the quantum mechanics associated to the D0-D2-D4-D6 realisation of the black hole \cite{Chowdhury:2015gbk,Chowdhury:2014yca}.
  
The aim of this paper is to construct explicit examples of micro-state geometries of four- dimensional BPS black holes.
  We follow the  Bena-Warner ansatz \cite{Bates:2003vx,Bena:2007kg,Gibbons:2013tqa}  and look for regular five-dimensional geometries generated by 
   distributions of half-BPS D-brane atoms  in $\mathbb{R}^3$.  The regularity of the five dimensional geometry is coded in the so called
    bubble equations that we generalise to account for the  case of branes at angles. Boundary conditions at infinity  further restrict the choices leading generically to a moduli space that consists of disjoint components. There are two classes of solutions. Scaling solutions are  configuration that can be rigidly scaled (see \cite{Denef:2002ru,Denef:2007vg,deBoer:2008zn,Bena:2012hf} for some results). The other class includes solutions where the distances between the centers are bounded by the charges.  
    
 The micro{-}state geometries that we find carry in general non-zero angular momentum. This may look surprising, since single{-}center BPS black holes in four dimensions cannot carry angular momentum, because rotations of a black hole horizon are not compatible with supersymmetry. 
  In our case, like for the  multi center BPS black holes with non zero angular momentum  considered in \cite{Elvang:2005sa}, the angular momentum is generated by the crossed electric and magnetic fields of  charges separated on $\mathbb{R}^3$. In the spirit of the fuzzball proposal, the black hole can be viewed as the superposition of  an  ensemble of geometries  with all  allowed angular momenta. The  statistical average exposes zero angular momentum, even though each micro-state can carry some\footnote{Here we have in mind a canonical ensemble interpretation. We thank Iosif Bena for clarifications on this point.}.

The plan of the paper is as follows. In Section \ref{BHfromInterD3} we present the BPS solutions describing systems of intersecting D3-branes on $T^6$ from the four dimensional perspective. We consider both cases of orthogonal and of intersecting D3-branes at angles.  The Bena-Warner ansatz is introduced in Section \ref{BenaWarnerEtc}. The ansatz is generalised to accomodate for non-orthogonally intersecting D3-branes.  In section  \ref{fuzzball1} and \ref{fuzzballangle}  solutions to the bubble equations and to the boundary conditions are found for the 3-center case with orthogonal or D3-branes intersecting at angles. A preliminary discussion of the counting of the number of micro-states with fixed charges is presented in the Conclusions. In Appendix \ref{appN1} we derive the four dimensional solution from dimensional reduction of the intersecting D3-branes solution in ten dimensional type IIB supergravity.

\section{Black holes from intersecting D3-branes }
\label{BHfromInterD3}

In this paper we consider a family of BPS solutions describing systems of intersecting D3-branes on $T^6$ from the four dimensional perspective. 
We refer the reader to the appendix for details on the ten dimensional solution and its reduction down to four dimensions. The four dimensional solution has been first explicitly derived in \cite{Bates:2003vx}  in the D0-D2-D4-D6 type IIA frame  and lifted to a system of M5 branes in M-theory. Here, for convenience, we consider the very symmetric formulation in terms of D3-branes, but the results can be easily translated into the type IIA and M-theory frame.   

The four-dimensional geometries can be viewed as solutions of an ${\cal N}=2$ truncation of  ${\cal N}=8$ supergravity involving the gravity multiplet and three vector multiplets. The  scalars $U^I$ in the vector multiplets,  usually referred as STU, parametrises the complex structures of the three internal $T^2$'s and span the moduli 
space  ${\cal M}_{STU}= [SL(2,R)/U(1)]^3 \subset  E_{7(+7)}/SU(8)= {\cal M}_{{\cal N} =8}$. Setting $16\pi G=1$, the lagrangian can be written as  
\eq
\lagr= \sqrt{g_4}\left( R_4-\sum_{I=1}^3 \frac{\partial_\m U_I\partial^\m \bar{U}_I}{2\pt{\Imm U_{I}}^2}-\frac{1}{4}F_a \mathcal{I}^{ab}F_{b}-\frac{1}{4}F_{a}\mathcal{R}^{ab} \widetilde{F}_{b} \right)
\feq
where
\begin{align}
 U_I=(\si+is,\tau+it,\nu+i u ) \qquad F_{a}=dA_a \qquad   \widetilde F_{a}=*_4 F_a
\end{align}
 are the three complex scalars in the vector multiplets characterising the complex structures of  the internal $T^2\times T^2\times T^2$. In these variables the kinetic functions read  
\begin{align*}
\mathcal{I}^{ab}&= stu \pmat 1+\frac{\si^2}{s^2}+\frac{\tau^2}{t^2}+\frac{\nu^2}{u^2} & \:\:{-}\frac{\si}{s^2}&\:\:{-}\frac{\tau}{t^2}&\:\:{-}\frac{\nu}{ u^2} \\
-\frac{\si}{s^2}  & \frac{1}{s^2} &0 &0 \\
-\frac{\tau}{t^2}  & 0 &\frac{1}{t^2}&0\\
-\frac{\nu}{ u^2}  & 0 & 0 &\frac{1}{u^2}\fpmat	 \quad	 
\mathcal{R}^{ab}=\pmat 2\si\tau\nu &\:\:{-}\tau\nu&\:\:{-}\nu \sigma& \:\:{-}\sigma\tau \\
-\tau\nu & 0 & \nu &\tau \\
-\nu \sigma  &\nu &0&\sigma\\
-\sigma\tau   &\tau &\sigma &0 \fpmat \num
\end{align*}
  The solutions  will be written in  terms of eight harmonic functions 
    \eq
  \{ V, L_I, K_I, M \} \label{harmh}
   \feq   
  on $\mathbb{R}^3$. It is convenient to introduce  the  combinations
\begin{align}\label{pzm}
Z_I&=L_I+\frac{|\ve_{IJK}|}{2}\frac{K_JK_K}{V},\nn\\
\m&=\frac{M}{2}+\frac{L_IK_I}{2V}+\frac{|\ve_{IJK}|}{6}\frac{K_IK_JK_K}{V^2}. 
\end{align}
 Here
  $\epsilon_{IJK}$ characterise the triple  intersections  among the three $T^2_I$ 2-cycles in $T^6$.
 
 The solutions can then be written as
\begin{align} \label{4dsolution}
ds^2&=-e^{2U}\pt{dt+w}^2+e^{-2U}\sum_{i=1}^3 dx^2_i,  \nn\\
A_a& =(A_0,A_I)=w_a +a_a\pt{dt+w}  \nn\\    
U_I &= -b_I+ i \pt{V e^{2U}Z_I}^{-1}
   \end{align}  
with
\begin{align}
b_I  &={K_I\over V} -\frac{\m}{Z_I}   \quad , \quad a_0=-\m V^2 e^{4U}  \quad , \quad a_I= V e^{4U} \, \left(  -{Z_1 Z_2 Z_3\over Z_I}  +K_I \, \m    \right)  \nn\\
% \al&=w_0-\m V^2 e^{4U}\pt{dt+w}     \quad\quad \al_I=-\frac{dt+w}{Z_I}+b_I  \,w_0+w_I \nn\\
   *_3dw_0&=dV   \quad , \quad    *_3 dw_I=-dK_I \quad , \quad
*_3dw=\half\pt{VdM-MdV+K_IdL_I-L_IdK_I}  \label{4dsolution0}
\end{align}
and
\bea
&&e^{-4U} =  {\cal I}_4(L_i,V,K_I,M)\equiv Z_1Z_2Z_3V-\m^2V^2 \\
&&=  L_1\, L_2 \, L_3 \, V- K_1\, K_2\, K_3\, M +\ft12 \sum_{I>J}^3   K_I K_J  L_I L_J  - {M V\over 2}  \sum_{I=1}^3 K_I L_I    -\ft14 M^2 V^2
-\ft14 \sum_{I=1}^3 K^2_I L_I^2  \nn
\eea

\subsection{The asymptotic geometry}

For general choices of the eight harmonic functions, the solution (\ref{4dsolution}) is singular. Both naked and `horizon-dressed' curvature singularities can be present. 
The generic solution is characterised by a mass $\mathfrak M$,   associated to the Killing vector $\xi^{(t)M} \partial_M= \partial_t$,  four  electric charges $Q_a$ and four magnetic charges $P_a$. Introducing the symplectic vector 
\bea
{\cal F}= \left(
\begin{array}{c}
  F_a  \\
 {\delta {\cal L}\over \delta F_a}   \\
\end{array}
\right) = \left(
\begin{array}{c}
  F_a  \\
  \star_4 \, I^{ab} \, F_b-R^{ab}\, F_b   \\
\end{array}
\right)
\eea
    one finds for the charges
    \bea
\mathfrak M &=& -\frac{1}{8\pi G}   \int_{  S^2_{\infty}  } \star_4 \,  d \xi^{(t)}    
     \nn\\
  \left(
\begin{array}{c}
  P_a  \\
  Q^a   \\
\end{array}
\right)    &=&      -\frac{1}{4 \pi}   \int_{  S^2_{\infty}  }  {\cal F}
%Q^a &=& -\frac{1}{4 \pi}   \int_{  S^2_{\infty}  }  \star_4\,   {\cal I}^{ab}\, F_b    \nn\\
%P_a  &=& -\frac{1}{4 \pi} \int_{  S^2_{\infty}  }   F_a    
\eea
 with  $\xi^{(t)} =\xi^{(t)}_M\, dx^M   $ and  $S^2_{\infty}$   the two sphere at  infinity.   Solutions with extra symmetries   arise for special choices of the harmonic functions. Axially symmetric solutions are characterised by the existence of an additional Killing vector $\xi^{(t)M} \partial_M= \partial_\phi$ associated to rotations around an axis in $\mathbb{R}^3$, and carry an extra quantum number, the angular momentum $J$ given by\footnote{Being a surface integral, the expression for $J$ holds true even when $\xi^{(\phi)}$ is only an asymptotic Killing vector.}
   \be
J = -\frac{1}{16 \pi G}   \int_{  S^2_{\infty}  } \star_4 \,  d \xi^{(\phi)}  \label{jj}
     \ee
  with  $\xi^{(\phi)} =\xi^{(\phi)}_M\, dx^M$. Spherical symmetric solutions are invariant under rotations around the origin and are characterized by  zero angular momentum. 

 In this paper we consider fuzzballs of spherically symmetric black holes.  The harmonic functions specifying the general spherically symmetric  solution can be written in the single-center form
   \begin{align}
 V=v_0+{v\over r}    \qquad     L_I=l_{0I}+{l_I\over r}  \qquad  K_I=k_{0I}+ {k_I\over r} \qquad M=m_0+{m\over r}  \label{bcH}
 \end{align}
  and describe a general system of intersecting D3-branes wrapping three cycles on $T^2\times T^2\times T^2$ with one leg on each of the three $T^2$.  
  The absence of Dirac-Misner strings requires that $w$ vanishes at infinity or, equivalently, that  $*_3dw \sim r^{-3} $ at infinity leading to the constraint 
  \be
  v_0 \, m-m_0 \, v+ k_{I0} \, l_I-l_{I0}\, k_I=0 \label{dmstring}
  \ee
   For simplicity  we take $m_0=m=0$. For this choice one finds
  \be
  e^{-4U}= V \, L_1\, L_2\, L_3-\ft14 \left( \sum_{I=1}^3 K_I L_I \right)^2 
  \ee
   Poles and zeros of this function are associated to horizons and curvature singularities respectively. If  $e^{-4U}>0$ for all $r>0$  the solution describes a black 
   hole with near horizon geometry $AdS_2\times S^2$ and entropy proportional to
 \be
 \lim_{r\to 0} r^4\, e^{-4U}  = {\cal I}_4  =  v \, l_1\, l_2\, l_3 -\ft14 \left( \sum_{I=1}^3 k_I l_I \right)^2>0
 \ee
    If $e^{-4U}$ has zeros for some positive positive $r$, the solution exposes a naked singularity.   

The charges of the solution (or its fuzzball) are computed by the integrals (\ref{charges}) evaluated in the  asymptotic geometries (\ref{bcH}). Writing the 
three-dimensional metric   in spherical coordinates
\be
ds^2=-e^{2U} (dt+w)^2+e^{-2U}\,  (dr^2+r^2 \, d\theta^2+r^2\,\sin^2\theta\, d\phi^2)
\ee
 and setting $G=(16\pi)^{-1}$ one finds for the charges\footnote{In our conventions $\star_4 \, dr\wedge dt=e^{-2U}\, r^2\, \sin\theta \, d\theta\wedge d\phi$ and 
 $\int \sin\theta d\theta \wedge d\phi  =4\pi$.}
    \bea
\mathfrak M &=&  8\pi  \, {r^2 \, \partial_r \, e^{2U}}  
     \nn\\   
\left(
\begin{array}{c}
  P_a  \\
  Q^a   \\
\end{array}
\right)    &=&        
\left(
\begin{array}{c}
  (v,-k_I)^T \\
    -r^2 \,  {\cal I}^{ab}\, \partial_r  a_b    \\
\end{array}
\right)   
   \label{charges}
\eea
   where we used the fact that at infinity $w=0$ and $F_a=dw_a+d a_a \, dt$.  
   On the other hand the angular momentum of the  fuzzball is computed by the integral (\ref{jj}). 
   We notice that the evaluation of this integral requires a more detailed knowledge of the asymptotic geometry since the angular momentum arises from the first dipole mode in the expansion of the harmonic function. Indeed, denoting
   \be
 H =h_0+  {h_1\over r} +{ \vec{h}_2 \cdot \vec x \over r^3}      
   \ee 
    one finds for the angular momentum
    \be
{\vec J} =4\pi\,    \left[ m_0 \,\vec{v}_2- v_0 \,{\vec m}_2 + l_{0I} \, \vec{k}_{2I} - k_{0I} \,\vec{l}_{2I}  \right]   \label{jjj}
   \ee    
    We anticipate here that apart from the scaling solutions, all fuzzball solutions we will find here carry a non-trivial angular momentum.  We observe that for orthogonal branes  angular momentum is carried by K-components (see Appendix \ref{appN1basic} for details)  corresponding to open string condensates on disks with boundary on two different D3-brane stacks.  Indeed, an explicit  microscopic  description  of the general supergravity solution  exists if the harmonic functions satisfy the boundary conditions  \cite{Bianchi:2016bgx}
    \be
     m_2+\sum k_{I2}=0   \label{micro}
    \ee    
   As we will see,  only scaling solutions in the list of examples we find satisfy this restriction on the dipole modes.

    \subsubsection{  Orthogonal branes} 
  
 We first consider the supergravity solution characterised by the harmonic functions
  \begin{align}
 V=1+{v\over r}    \qquad     L_I= 1+{l_I\over r}  \qquad  K_I= M= 0  \label{bcH1}
 \end{align} 
 describing a system of four stacks of D3-branes intersecting orthogonally on $T^6$.  At large distances one finds
 \bea
 e^{-2U} &=& \sqrt{VL_1L_2L_3}=1+{( v+l_1+l_2+l_3)\over 2\,r}  +\ldots   \nn\\
 a_I &=& -L_I^{-1}=-1+{\l_I\over r}+\ldots    \qquad   a_0=0 \nn\\ 
 U_I &=&  i \pt{V e^{2U} L_I}^{-1} = i+\ldots 
 \eea
  leading to 
    \bea\label{massort}
\mathfrak M &=& 4\pi\, ( v+l_1+l_2+l_3) 
     \\
    \label{chargeort} \left(
\begin{array}{c}
  P_a  \\
  Q^a   \\
\end{array}
\right)    &=&        
\left(
\begin{array}{c}
  (v, 0,0,0)^T \\
   (0,l_1,l_2,l_3)^T  \\
\end{array}
\right)   
\eea
  The extremal Reissner Nordstrom solution corresponds to the choice $l_I=v=Q/2$, or equivalently 
  \be
  L_I=V=1+{l \over  r}   \qquad M=K_I=0
  \ee 
   after the identification  $ F_{\rm RN}=\ft12 \left(   * F_0+\sum_{I=1}^3 F_I  \right)$.

    \subsubsection{ Branes at angles}
   
    We next consider the supergravity solution characterised by the harmonic functions
     \begin{align}
 V=1+{v\over r}    \qquad     L_I=1+{l_I\over r}  \qquad  K_1=g +{ k_1\over r}  \qquad K_2=g\qquad K_3=M=0 \label{bcH2}
 \end{align}
   The absence of Dirac-Misner strings (\ref{dmstring}) requires $k_1=g(l_1+l_2)$.  The resulting solution is equivalent after a duality transformation to  the solution found in \cite{Bertolini:2000ei} describing a system of D3 branes intersecting at a non-trivial angle between the branes at infinity, parametrised by $g$. 
   
    The asymptotic solution at large $r$ becomes
      \begin{align}\label{pzm2}
        e^{2U} &= 1-{v+l_1+l_2+l_3 \over 2 r}+\ldots \qquad  U_I  = \left({\rm i} , {\rm i} ,    {  1\over g-{\rm i}  }  \right)  +  \ldots  \nn\\
         w_a &= ( v\,\cos\theta\,d\phi  ,-k_1\,\cos\theta\,d\phi  ,0,0)\,   +\ldots \nn\\
         a_a &=   \left( {g\, l_3 \over r}  , {(1+g^2) l_1\over r}  ,  {l_2-g^2 l_1\over r}  ,{l_3\over r}  \right) +\ldots \nn\\
      {\cal I}^{ab}&=   \left(
\begin{array}{cccc}
 1 & 0 & 0 & -g \\
 0 & \frac{1}{1+g^2} & 0 & 0 \\
 0 & 0 & \frac{1}{1+g^2} & 0 \\
 -g & 0 & 0 & 1+g^2
\end{array}
\right)   \qquad     {\cal R}^{ab}=   \left(
\begin{array}{cccc}
 0 & 0 & 0 & 0 \\
 0 &  0  &  \frac{g}{1+g^2}  & 0 \\
 0 &  \frac{g}{1+g^2}  & 0 & 0 \\
 0 & 0 & 0 & 0
\end{array}
\right)
 \end{align}
 with dots denoting higher order terms in the  expansion for large $r$.  Plugging  (\ref{pzm2}) into (\ref{charges}) one finds for the charges\footnote{For the central charges $Z_I=e^{-U}\,    (2 {\rm Im}U_I)^{-1} \, r^2 \, \partial_r U_I$,  $Z_4={\rm i}\, e^{-U}\, r^2\,\partial_r U$ one finds at infinity
 \bea
  Z_1 &=&4\pi\left[   2 \,g \,l_1+{\rm i} (v+l_1-l_2-l_3) \right] \qquad Z_2  =4\pi \left[ - 2 \,g\, l_1+{\rm i} (v-l_1+l_2-l_3 ) \right] \nn\\
  Z_3 &=& 4\pi  {\rm i} (v-l_1-l_2+l_3) \qquad Z_4  = 4\pi {\rm i}  (v+l_1+l_2+l_3 ) 
 \eea
 showing the saturation of the BPS bound $\mathfrak{M}=|Z_4|\ge |Z_i|$ for $i\neq 4$, when $v$ and $l_I$ all have the same sign and $g$ is sufficiently small.
}
      \bea
\label{massang}\mathfrak M &=& 4\pi\, ( v+l_1+l_2+l_3) 
     \\
\label{chargeang}     \left(
\begin{array}{c}
  P_a  \\
  Q^a   \\
\end{array}
\right)    &=&        
\left(
\begin{array}{c}
  (v, -g\, (l_1+l_2) ,0,0)^T \\
   (0,l_1,l_2,l_3)^T  \\
\end{array}
\right)   
\eea
    Due to different conventions and duality frames, it is not easy to compare the charges with the corresponding ones in \cite{Bertolini:2000ei}.
 \section{ Microstate geometries}
 \label{BenaWarnerEtc}
 
 In this section we review the Bena-Warner  multi-Taub NUT ansatz for fuzzball geometries of four- and five-dimensional black holes. We slightly 
 generalise the ansatz to accomodate for non-orthogonal brane intersections and derive the corresponding bubble equations. In the next section we present explicit horizon-free solutions with three centers.
 
 \subsection{The eleven dimensional lift}

The four-dimensional solution (\ref{4dsolution})  lifts to an eleven dimensional solution representing a systems of intersecting M5-branes with four electric and four magnetic charges. The eleven dimensional metric is given by  \cite{Bena:2007kg}:
\bea
ds^2=ds^2_5+ds^2_{T^6}  \label{11d}
\eea
where
\bea
ds_5^2 &=& -(Z_1\, Z_2\, Z_3)^{-{2\over 3}} \left[dt+\mu(d\Psi+w_0) +w\right]^2 + (Z_1\, Z_2\, Z_3)^{1\over 3}\left[ V^{-1}  (d\Psi+w_0)^2+ V d\vec x^2 \right] \nn\\
ds_{T^6} &=& \sum_{I=1}^3 \left( { Z_1\, Z_2\, Z_3 \over Z_I^3 } \right)^{ 1\over 3} (dy_I^2+d\tilde y_I^2) 
\eea
in which the coordinates associated to $\mathbb{R}\times S^1 \times T^6$ are respectively $\lbrace t, \, \vec x ,\, \Psi, \, y_I, \,\tilde y_I \rbrace$ with $I=1,2,3$.
  Micro{-}states of the four dimensional black holes can be generically defined as smooth geometries with no horizons or curvature singularities in eleven dimensions carrying the same mass and charges as the corresponding black hole.  Regular solutions can be constructed in terms of multi-center harmonic functions $(V,L_I,K_I,M)$ with the positions of the centers and the charges chosen such that  $Z_I$ are finite and $\mu =0$ near the centers. Under these assumptions one finds that the eleven dimensional metric (\ref{11d})  near the centers is $\mathbb{R}\times T^6\times \mathbb{R^4}/\mathbb{Z}_{|q_i|}$. To avoid orbifold singularities we will henceforth take $|q_i|=1$.  
  Moreover, the  absence of horizons and closed time{-}like curves  requires that
    \bea
  && Z_I V >0 \qquad {\rm and} \qquad   e^{2U}>0
  \eea
   Let us remark that the  condition $Z_I  V>0$  near the centers requires 
  \be
\left.Z_I\,V\right|_{r_i=0}=q_i\,\left(l_{0I}+\sum_{j\ne i}\frac{l_{I,j}}{r_{ij}}\right)+l_{I,i}\left(v_0+\sum_{j\ne i}\frac{q_j}{r_{ij}}\right)+C_{IJK} k^J_i\left( k_{0}^K+\sum_{j\ne i}\frac{k_j^K }{r_{ij}}\right) >0
  \ee
It turns out that these necessary conditions often are enough to ensure the positivity of both $Z_I V$ and $e^{2U}$ on the whole $\mathbb{R}^3$. 
  In the next section we look for explicit  solutions of these requirements satisfying the boundary conditions (\ref{bcH}). We stress that the resulting solutions are regular everywhere in five dimensions and fall off at infinity to $\mathbb{R}^{1,3}\times S^1$. The four-dimensional fuzzball solution  follows from reduction of this five-dimensional geometry down to four dimensions where the apparent singularity in the geometry is balanced by a blow up of the scalar fields.

  \subsection{The  bubble equations} 
 
  We consider N-center harmonic functions following the ansatz
 \bea
 V &=& v_0+\sum_{i=1}^N  {q_i\over r_i }  \qquad  ,  \qquad      L_I = l_{0I}+ \sum_{i=1}^N   { l_{I,i} \over  r_i }  \nn\\ 
 K^I &=& k^I_{0}+\sum_{i=1}^N  {k^I_{i}\over r_i }  \qquad  ,  \qquad M =  m_0+\sum_{i=1}^N   { m_i \over  r_i }  \label{ansatz0}
 \eea
 with $ r_i =| {\bf  y}_i -{\bf x}|$  and  ${\bf y}_i$ the position of the i$^{\rm th}$ center. We notice that $(\ell_{Ii} , m_i)$ and $(q_i,k^I_i)$ describe the electric and 
 magnetic  fluxes of the four dimensional gauge fields through the sphere encircling  the i$^{\rm th}$- centers, so Dirac quantisation requires that  
 they be quantised.  Here we adopt units such that they are all integers. 
 Alternatively, one can think of the eight charges as parametrising the number of D3-branes wrapping one of the eight three{-}cycles with a leg on each of the three tori $T^2_I$ in the factorisation $T^6 = T^2_1 \times T^2_2\times T^2_3$. In other words, in our units  each charge describes the number of D3-branes of a certain kind.
 
 We look for regular five dimensional geometries behaving as $\mathbb{R}\times {\rm Taub{-}NUT}$ near the centers. It is easy to see that $w$ vanishes near the centers, so the Taub-NUT geometry factorises if $Z_I$ are finite and $\mu$ vanishes near the centers, i.e. 
  \bea
 Z_I\big|_{r_i\approx 0}  &\approx&  \zeta^I_i    \nn\\
      \mu  \big|_{r_i \approx 0} &\approx&  0
 \eea
  with $\zeta^I_i$ some finite constants.  The conditions that $Z_I$ is finite near the centers  can be solved by taking
 \bea
  \ell_{I,i} &=& -   {|\epsilon_{IJK}|\over 2}  {k^J_{i} \, k_i^K\over q_i  }   \nn\\
    m_i  &=&    {k^1_i \, k^2_{i} \, k_i^3\over q_i^2 } \label{halfbps} 
 \eea
The vanishing of $\mu$ near the centers boils down to the so called {\it bubble equations}
 \bea \label{eq: bubble equations}
 \sum_{j=1}^N \Pi^{(1)}_{ij}\, \Pi^{(2)}_{ij} \, \Pi^{(3)}_{ij}\, {q_i \,q_j\over  r_{ij} }+v_0 \,{k^1_i\, k^2_i\, k^3_i \over q_i^2 } -\sum_{I=1}^3 l_{0I}\, k^I_i  -  
|\epsilon_{IJK}| \, {  k_{0I} \, k_i^J \, k_i^K \over 2\,q_i} -m_0 q_i=0
 \eea
 with
 \be
  \Pi^{(I)}_{ij}={k^I_i\over q_i}-{k^I_j\over q_j}  \qquad   r_{ij}=| {\bf y}_i-{\bf y}_j|
 \ee
  Indeed the conditions (\ref{halfbps}) ensure that both $Z_I$ and $\mu$ are finite near the centers while the bubble equations follows from the requirement
  that $\mu$ vanishes near the center. The bubble equations ensure also the absence of Dirac-Misner strings. To see this, we notice that using 
  the bubble equations, the $w$ function  defined by (\ref{4dsolution0}) can be written in the form
   
   \bea
   *_3 dw &=&  \sum_{i,j=1}^N ( q_{[i} \, m_{j]}   +k^I_{[i} \, l_{j],I }   )  {1\over r_i}d  {1\over r_j } +\half\sum_{i=1}^N  \left( v_0\, m_i-m_0 \, q_i-\sum_{I=1}^3 ( l_{0I}\, k_i^I-
    k_{I0} \, l_{i,I} )  \right)\, d {1\over r_i }  \nn \\  
  %  &=& \half\sum_{i,j=1}^N\, \Pi^{(1)}_{ij}\, \Pi^{(2)}_{ij} \, \Pi^{(3)}_{ij}\, q_i \,q_j \,  {1\over r_j}d  {1\over r_i } 
  %   +\half \sum_{i=1}^N \left( {k^1_i\, k^2_i\, k^3_i \over q_i^2 } -\sum_{I=1}^3 k^I_i  \right)\,  d {1\over r_i } \nn\\
     &=&  \half\sum_{i,j=1}^N\, \Pi^{(1)}_{ij}\, \Pi^{(2)}_{ij} \, \Pi^{(3)}_{ij}\, q_i \,q_j \,\left(   {1\over r_j} -{1\over r_{ij} }  \right)  d{1\over r_i}   \label{dwdm}
   \eea
    where in the second line we used equations (\ref{halfbps}), (\ref{eq: bubble equations}) and $A_{[BC]}$ means $\frac{1}{2}(A_{BC}-A_{CB})$. The solution can be written in the form
      \bea
   w  &=& \quar\sum_{i,j=1}^N\, \Pi^{(1)}_{ij}\, \Pi^{(2)}_{ij} \, \Pi^{(3)}_{ij}\, q_i \,q_j \,\omega_{ij}   \label{wiwij}
   \eea 
    in terms of the one forms $\omega_{ij}$ defined via the relation
\eq 
*_3 d\om_{ij}=\pt{\frac{1}{r_j}-\frac{1}{r_{ij}}}d\frac{1}{r_i}-\pt{\frac{1}{r_i}-\frac{1}{r_{ij}}}d\frac{1}{r_j},
\feq
that is
     \be
   \omega_{ij} =\frac{\pt{\mbf{n}_i+\mbf{n}_{ij}}\cdot\pt{\mbf{n}_j-\mbf{n}_{ij}}}{ r_{ij} }d\phi_{ij}   \label{wij}
   \ee
with
 \be
       {\bf n}_i ={{\bf x}- {\bf y}_i   \over r_i}   \qquad      {\bf n}_{ij} ={ {\bf y}_i -{\bf y}_j \over r_{ij} } \nn\\
   \qquad 
 d{\phi}_{ij} =  \frac{ {\bf n}_{ij}\times {\bf n}_i \cdot dx}  {    r_i  \left[ 1- ({\bf n}_{ij} \cdot {\bf n}_i)^2 \right]   }.  
\ee
  It is easy to see that $\omega_{ij}$ is free of Dirac-Misner singularities. Indeed along the dangerous lines connecting any two centers the numerator of  (\ref{wij}) always vanish so no string{-}like singularity arises. One can also see that near the centers $w$ goes to a constant and exact form. 
 
  Finally we notice that  if the coefficients $k^I_i$ satisfy the  relation
   \be
   v_0 \, m_i -\sum_{I=1}^3 l_{0I}\, k^I_i  +  
 k_{0I} \,  l_{Ii} -m_0 q_i=0
    \label{scaling}
   \ee
   the system of equations is invariant under  overall rescalings of the center positions ${\bf y}_i \to \lambda {\bf y}_i$. These solutions are known as ``scaling solutions".  Multiplying equation (\ref{scaling}) by the positions of the centers ${\vec y}_i$ and summing one finds that  the scaling solutions satisfy 
   \be
    m_0 \,\vec{v}_2- v_0 \,{\vec m}_2 + l_{0I} \, \vec{k}_{2I} - k_{0I} \,\vec{l}_{2I} =0
   \ee    
   and therefore according to (\ref{jjj}) they carry zero angular momentum.

 \section{ Fuzzballs of orthogonally intersecting branes}
 \label{fuzzball1}
 
  We look for regular geometries with the asymptotics (\ref{bcH1}), i.e.   
  \be
  l_{0I}=v_0=1  \qquad  m_0=m=k_{0I}=k_I=0
  \ee
  For concreteness we take $q_i=1$. 
   The  charges  of the fuzzball solutions are then
  \bea
P_0 &=& N \nn\\
Q_I &=&    -    \sum_{i=1}^N \, {   |\epsilon_{IJK}|   k^J_{i} \, k_i^K \over 2}  \label{chargesortho}
 \eea 
    The  solution is specified by the  the positions ${\bf y}_i$ of the centers and the fluxes  $k_i^I$. The positions of the centers are constrained by the bubble equations
  \bea
       &&  \sum_{j \neq i}^N  { k^{(1)}_{ij } \, k^{(2)}_{i j}  \,  k^{(3)}_{i j}    \over  r_{ij} } + k^1_i\, k^2_i\, k^3_i  -\sum_{I=1}^3 k^I_i    =0   
                       \label{eqsf}
   \eea
    with $k_{ij}^{(I)}=k_i^{I}- k_j^{I}$ while the match of the asymptotic geometries requires
    \bea
         \sum_{i=1}^N \,k_i^I=   \sum_{i=1}^N \,   k^1_i \, k^2_{i} \, k_i^3  =0 \label{bcasymp}
    \eea
     In addition, the absence of horizons and of closed time-like curves requires
  \be
  Z_I\, V>0  \qquad  {\rm and} \qquad    e^{2U} >0
  \ee
    We will consider solutions with three centers. Configurations with one or two centers fail to meet the requirement $Q_I>0$. 
       
\subsection{Three centers}

The bubble equations  (\ref{eqsf})  for three centers can be solved in general by taking
\be
r_{12}=\frac{\Pi_{12}\,r_{23}}{\Pi_{23}-r_{23}\left(\Gamma_2 -\Lambda_2\right)}\qquad\qquad r_{13}= \frac{\Pi_{13}\,r_{23}}{-\Pi_{23}+ r_{23}\left(\Gamma_1+\Gamma_2-\Lambda_1-\Lambda_2 \right)}     \label{rij3}
\ee
with
\be
\Pi_{ij}=\prod_{I=1}^3 (k^I_i -k^I_j)  \qquad\qquad \Gamma_i=\sum_{I=1}^3 k^I_{\ i}\qquad\qquad \Lambda_i=k^1_{\ i}k^2_{\ i}k^3_{\ i}
\ee
%The boundary conditions (\ref{bcasymp}) require that  $k^I {}_i$ takes the form 
% \be
% k^I {}_i=
%\left(
%\begin{array}{ccc}
% k_1 & k_2   & -k_1-k_2  \\
% k_3 & k_4   & -k_3-k_4 \\
% k_5 & k_6   & -k_1-k_2  
%\end{array}   \label{charges3}
%\right)  
% \ee 
%with
% \be
%  k_1=-\frac{k_2(k_4k_5+k_3k_5+k_3k_6)}{k_3k_6+k_4k_5+k_4k_6}.  \label{charges3B}
% \ee 
% Plugging (\ref{charges3}) and (\ref{charges3B}) into (\ref{chargesortho}) 
% one finds for the charges
%\begin{align*}
%Q_1&=-k_3(2k_5+k_6)-k_4(k_5+2k_6)\\
%Q_2&=k_2\frac{k_4(k_5^2-2k_5k_6-2k_6^2)+k_3(2k_5^2+2k_5k_6-k_6^2)}{k_3k_6+k_4k_5+k_4k_6}\\
%Q_3&=k_2\frac{k_3^2(2k_5+k_6)+2k_3k_4(k_5-k_6)-k_4^2(k_5+2k_6)}{k_3k_6+k_4k_5+k_4k_6}\num.
%\end{align*}
  A solution given by (\ref{rij3}) makes sense if the distances $r_{ij}$ between the three centers are positive and they satisfy the triangle inequalities.  This  restricts significantly the choices for the $k_i^I$. A quick scan over the integers shows that boundary conditions are solved only if at least one of the fluxes $k_i^I$ vanishes.
  Without loss of generality the general solution can then be parametrised in the form (up to permutations of rows and columns)
   \bea
 k^I {}_i &=&
\left(
\begin{array}{ccc}
-\kappa_1\,\kappa_2 & -\kappa_1\,\kappa_3   & \kappa_1\,(\kappa_2+\kappa_3)  \\
\kappa_3 & \kappa_2  & -\kappa_2-\kappa_3  \\
 -\kappa_4&  \kappa_4 &  0
\end{array}
\right)  
\eea
 Consequently the harmonic functions takes the general form
 \bea
V &=& 1+\sum_{i=1}^3 {1\over r_i}   \qquad   M= \kappa_1 \kappa_2\kappa_3\kappa_4 \, \left( { 1\over r_1}-{1\over r_2}\right) 
\nn\\
 L_1 &=&1+  \kappa_4 \left( {\kappa_3 \over r_1}-{\kappa_2\over r_2} \right)  \qquad  L_2 =1+  \kappa_1 \kappa_4 \left(  -{\kappa_2\over r_1}+{\kappa_3\over r_2} \right) \nn \\
 L_3 &=&1+  \kappa_1 \left( {\kappa_2\kappa_3 \over r_1}+{\kappa_2 \kappa_3\over r_2} +{(\kappa_2+\kappa_3)^2 \over r_3} \right) \qquad
K_1 =  \kappa_1 \left( -{\kappa_2 \over r_1}-{\kappa_3\over r_2}+{\kappa_2+\kappa_3\over r_3} \right) \nn\\
  K_2 &=&  {\kappa_3 \over r_1}+{\kappa_2 \over r_2}-{\kappa_2+\kappa_3 \over r_3}  \qquad 
  K_3 =  \kappa_4 \left( -{1\over r_1}+{1\over r_2} \right)  \nn
 \eea
 The charges and distances between the centers reduce to
 \bea
Q_1&=& \kappa_4(\kappa_3-\kappa_2)\qquad  Q_2= \kappa_1\kappa_4(\kappa_3-\kappa_2)\qquad  Q_3= \kappa_1(\kappa_2^2+4\kappa_2\kappa_3+\kappa_3^2) \nn\\
r_{12}&=& \frac{2\kappa_1\kappa_4(\kappa_2-\kappa_3)^2r_{23}}{\kappa_1\kappa_4(2\kappa_2^2+5\kappa_2\kappa_3+2\kappa_3^2)+(\kappa_2+\kappa_4-\kappa_1\kappa_3+\kappa_1\kappa_2\kappa_3\kappa_4)r_{23}}\nn\\
r_{13}&=& \frac{\kappa_1\kappa_4(2\kappa_2+\kappa_3)(\kappa_2+2\kappa_3)r_{23}}{\kappa_1\kappa_4(2\kappa_2^2+5\kappa_2\kappa_3+2\kappa_3^2)-(\kappa_1-1)(\kappa_2+\kappa_3)r_{23}}.
\eea

\subsubsection{Scaling solutions }

The scaling solution corresponds to the choice 
\be
\kappa_2=0  \qquad    \kappa_1=1  \qquad \kappa_3=\kappa_4=\kappa
\ee
One finds
 \bea
 k^I {}_i &=&
\left(
\begin{array}{ccc}
 0 & -\kappa   & \kappa  \\
 \kappa & 0  & -\kappa  \\
  -\kappa & \kappa  &   0
\end{array}
\right)   \qquad   r_{12}=r_{23}=r_{13}=\ell  \nn\\
P_0 &=& 3  \qquad ~~~~~~~~~ Q_1= Q_2=Q_3=\kappa^2    
 \label{kk}
 \eea
for any given $\ell$.   The regularity conditions  become  
  \begin{align}
& e^{-4U} =1+\frac{r_1 r_2+r_1
   r_3+r_2 r_3}{r_1 r_2 r_3}+\frac{\kappa^2 \left(r_1 r_2+r_1 r_3+r_2r_3+3 r_1+3 r_2+3 r_3\right)}{r_1 r_2 r_3}\nn\\
&+\frac{\kappa^4 \left(r_1+r_2+r_3+9\right)}{r_1 r_2
   r_3}  +\frac{\kappa^6 \left(2 r_1 r_2+2 r_1 r_2+2 r_2r_3+ r_1 r_2 r_3-r_1^2 -r_2^2-r_3^2\right)}{r_1^2 r_2^2
   r_3^2}>0\nn\\
   Z_1V &=1+\frac{ r_1 \,r_2 +r_1\, r_3+ r_2\, r_3 + \kappa^2 (2 \, r_2 +2 r_3-r_1+r_2\, r_3  ) }{r_1r_2r_3}>0
\end{align}
 The conditions  $Z_2 V>0$ and  $Z_3 V>0$ follow from  $Z_1V >0$ and the permutation symmetry of the system. 
The two conditions can be shown to be satisfied using  the inequality
 \be
r_1+r_2-r_3 \geq 0
 \ee
 holding for any point $x\in \mathbb{R}^3$ if  $r_i$ denotes the distance from the point to the three vertices of an equilateral triangle. This inequality can be proved using triangle inequalities \cite{iannuzzitrapani}.    
  
   We conclude that the five-dimensional geometry defined by the multi-center solution is regular everywhere. We notice that the  fluxes satisfy the scaling condition
   (\ref{scaling}) and consistently a rigid rescaling of the positions of the centers generate a new solution.  More precisely, the moduli space of solutions with this charge is spanned by a single continuous parameter $\ell$ and permutations of the rows or columns of the matrix (\ref{kk}). 
    There are 12  inequivalent choices corresponding to the $3!$ permutations of the entries in the first line in (\ref{kk}) times the two choices for the position of the 0 in the second line. The remaining entries are determined by the conditions that the sum along rows and columns of the matrix $k_i^I$ should vanish. 
   The number $12$ matches the number of apostles and  the degeneracy of  four-charge black hole micro{-}states with the minimal unit of charge  $P_0=Q_I=1$ \cite{Chowdhury:2015gbk}!    Moreover the solutions satisfy $\sum_{I=1}^3 \vec{k}^I_{2}=\vec{m}_2=0$ and   
therefore according to (\ref{jjj}) and (\ref{micro}) they carry zero angular momentum and admit a microscopic description in terms of orthogonal intersecting D3-branes along the lines of \cite{Bianchi:2016bgx}. 

This solution can been shown\footnote{We thank Iosif Bena for pointing this out to us.} to be related by dualities to a known supertube solution originally found in \cite{Vasilakis:2011ki}.

\subsubsection{Non-scaling solutions}

The analysis  above can be repeated for more general choices of the fluxes but  regularity conditions in general can only be verified numerically.  
Here we list some illustrative  examples of the type of solutions one finds.  

\begin{itemize}

\item{ $\kappa_2=0$,  $\kappa_1=\kappa_3=1$, $\kappa_4=\kappa:$
 \bea
 k^I {}_i &=&
\left(
\begin{array}{ccc}
0 & -1   & 1  \\
 1 & 0  & -1  \\
-\kappa&  \kappa &  0
\end{array}
\right)   \qquad   r_{13}=r_{23}\qquad r_{12}=\frac{2\, \kappa\, r_{23}}{2\,\kappa +(\kappa-1)\,r_{23}} \nn\\
P_0&=& 3 \qquad Q_1=Q_2=\kappa\qquad Q_3=1  
\eea
 Interestingly, triangle inequalities in this case do not constrain $r_{23}$ that can take arbitrarily large value. 
 
 }
 
  \item {$\kappa_2=\kappa_4=\kappa$, $\kappa_1=1$,\  $\kappa_3=2\,\kappa$ 
  \bea
 k^I {}_i &=&
\left(
\begin{array}{ccc}
-\kappa & -2\,\kappa   & 3\,\kappa \\
2\kappa & \kappa  & -3\,\kappa \\
 -\kappa&  \kappa &  0
\end{array}
\right)   \qquad   r_{12}=\frac{r_{23}}{10 + r_{23}}\qquad r_{13}=r_{23}.\nn\\
P_0&=& 3 \qquad Q_1=Q_2=\kappa^2\qquad Q_3=13\,\kappa^2  
\eea
 As before $r_{23}$  can take arbitrarily large value. 
}

 \item{ $\kappa_2=0$, $\kappa_1=3\,\kappa$,\  $\kappa_3=2\,\kappa$,   $\kappa_4=\kappa$
  \bea
 k^I {}_i &=&
\left(
\begin{array}{ccc}
0 & -3\,\kappa   & 3\,\kappa  \\
 \kappa & 0  & -\kappa  \\
-2 \kappa&  2\kappa &  0
\end{array}
\right)   \qquad   r_{12}=\frac{12\, \kappa^2\, r_{23}}{12\,\kappa^2 -r_{23}} \qquad r_{13}=\frac{6\, \kappa^2\, r_{23}}{6\,\kappa^2 -r_{23}}  \nn\\
P_0&=& 3 \qquad Q_1=2\,\kappa^2\qquad Q_2=6\,\kappa^2\qquad Q_3=3\,\kappa^2  \nn\\
r_{23} &<& 6\,(2-\sqrt{2})\,\kappa^2 
\eea
 We notice that triangle inequality in this case impose an upper bound on $r_{23}$ leading to a moduli space of finite volume.
}

\end{itemize}

 \section{ Fuzzballs of branes at angles}
  \label{fuzzballangle}
  
 We look for regular fuzzball geometries with the asymptotics (\ref{bcH2}), i.e. 
  \be
  l_{0I}=v_0=1  \qquad  m_0=m=k_{03}=k_3=k_2=0\qquad k_{01}=k_{02}=g\qquad k_1=g(l_1+l_2)
  \ee
   This describes a fuzzball of a non-rotating black hole with charges  
  \bea
P_0 &=& N \nn\\
Q_I &=&    -   {|\epsilon_{IJK}|\over 2}  \sum_{i=1}^N \, {k^J_{i} \, k_i^K\over q_i  }  
 \eea 
    The fuzzball is specified by the parameters  $k_i^I$  describing the magnetic fluxes through the two-spheres encircling the centers and
    the positions  ${\bf y}_i$. The positions of the centers are constrained by the bubble equations
  \bea
       &&  \sum_{j \neq i}^N  { k^{(1)}_{ij } \, k^{(2)}_{i j}  \,  k^{(3)}_{i j}    \over  r_{ij} } + k^1_i\, k^2_i\, k^3_i  -\sum_{I=1}^3 k^I_i - g\,k^2_i\,k^3_i-g\,k^1_i\,k^3_i =0    \label{eqsfg}
   \eea
    while the match of the asymptotic geometries requires
    \bea
         \sum_{i=1}^N \,k_i^2= \sum_{i=1}^N \,k_i^3=   \sum_{i=1}^N \,   k^1_i \, k^2_{i} \, k_i^3  =0\qquad \sum_{i=1}^N \,k_i^1=g\,(Q_1+Q_2)                 \label{bcasympg}
    \eea
     In addition, the absence of horizons and of closed time-like curves requires
  \be
  Z_I\, V>0  \qquad  {\rm and} \qquad    e^{2U} >0
  \ee

\subsection{Three centers}

The bubble equations  (\ref{eqsfg})  for three centers can be solved in general by taking
\be
r_{12}=\frac{\Pi_{12}\,r_{23}}{\Pi_{23}-r_{23}\left(\Gamma_2 -\Lambda_2+\Om_2\right)}\qquad r_{13}=-\frac{\Pi_{13}\,r_{23}}{\Pi_{23}- r_{23}\left(\Gamma_1+\Gamma_2-\Lambda_1-\Lambda_2+\Om_1+\Om_2 \right)}
\ee
with
\be
\Pi_{ij}=\prod_{I=1}^3 ( k^{I}_i-k^I_j)  \qquad \Gamma_i=\sum_{I}k^I_{\ i}\qquad \Lambda_i=k^1_{\ i}k^2_{\ i}k^3_{\ i}\qquad \Om_i=g\,k^2_i\,k^3_i+g\,k^1_i\,k^3_i
\ee
%The boundary conditions (\ref{bcasympg}) require that  $k^I {}_i$ takes the form 
% \be
% k^I {}_i=
%\left(
%\begin{array}{ccc}
% k_1 & k_2   & k_7  \\
% k_3 & k_4   & -k_3-k_4 \\
% k_5 & k_6   & -k_1-k_2  
%\end{array}  
%\right)  
% \ee 
%with ( la formula e' troppo lunga!!!!! )
% \bea
%  k_1&=&-\frac{g\,(k_3+k_4)(k_5+k_6)[k_3(2\,k_5+k_6)+k_4\,(k_5+2\,k_6)]+k_2[k_3(k_5+k_6)(1+g\,k_6)+k_4(k_5(1+2\,g\,k_6)+2\,g\,k_6^2)]}{k_3(k_6(1+2\,g\,k_5)+2\,g\,k_5^2)+k_4(k_5+k_6)(1+g\,k_5)}\nn\\
%k_7&=&\frac{g\,k_3\,k_5[k_3(2\,k_5+k_6)+k_4(k_5+2\,k_6)]+k_2[k_3\,k_5(1+g\,k_6)-k_4\,k_6(1+g\,k_5)]}{k_3(k_6(1+2\,g\,k_5)+2\,g\,k_5^2)+k_4(k_5+k_6)(1+g\,k_5)}
% \eea
% Plugging ...
% one finds for the charges ( le formule sono ancora un po' lunghe, non so se vale la pena scriverle.....)
%\begin{align*}
%Q_1&=\\
%Q_2&=\\
%Q_3&=\num.
%\end{align*}

We consider the case
 \be
 k^I {}_i=
\left(
\begin{array}{ccc}
 0 & -\kappa_1   &  \kappa_1 + g\,\kappa_3(\kappa_1+\kappa_2)  \\
\kappa_2 & 0   & -\kappa_2\\
-\kappa_3 & \kappa_3   & 0  
\end{array}  
\right)  
 \ee 
with $\kappa_1,\kappa_2,\kappa_3$ three positive integers and $g$ a rational number. 
 One finds
\bea
Q_1&=&\kappa_2\,\kappa_3\qquad Q_2=\kappa_1\,\kappa_3\qquad Q_3=\kappa_1\,\kappa_2+g\,\kappa_2\,\kappa_3(\kappa_1\,+\kappa_2 )\nn\\
r_{12}&=& \frac{2\,\kappa_1\,\kappa_2\,\kappa_3\,r_{23}}{2\,\kappa_1\,\kappa_2\,\kappa_3+g\,\kappa_2\,\kappa_3^2\,(\kappa_1+\kappa_2)-(\kappa_1-\kappa_3+g\,\kappa_1\,\kappa_3)\,r_{23}}\nn\\
r_{13}&=& \frac{2\,[\,\kappa_1\,\kappa_2\,\kappa_3+g\,\kappa_2\,\kappa_3^2\,(\kappa_1+\kappa_2)]\,r_{23}}{2\,\kappa_1\,\kappa_2\,\kappa_3+g\,\kappa_2\,\kappa_3^2\,(\kappa_1+\kappa_2)-[\kappa_1-\kappa_2+g\,\kappa_3\,(\kappa_1+\kappa_2)]\,r_{23}}
\eea
The harmonic functions become
\bea
V &=& 1+\sum_{i=1}^3 {1\over r_i}   \qquad    L_I=1+ {Q_I\over r_I}  \qquad   M=0 \\
K_1 &=&  g+\left( -{\kappa_1 \over r_2}+{\kappa_1+  k_1 \over r_3} \right)  \qquad  K_2 = g+ \kappa_2\left( {1\over r_1}-{1\over r_3} \right)\qquad 
K_3 =  \kappa_3 \left(- {1\over r_1}+{1\over r_2} \right)  \nn
 \eea

 Some examples are
 
 \begin{itemize}

\item{ $\kappa_1=\kappa_2=\kappa_3=\kappa=(2g)^{-1}$
 
 \bea
 k^I {}_i &=&
\left(
\begin{array}{ccc}
0 & -\kappa   & 2\kappa  \\
 \kappa & 0  & -\kappa  \\
 -\kappa&  \kappa &  0
\end{array}
\right)   \qquad   r_{12}=\frac{4\, \kappa^2\, r_{23}}{6\,\kappa^2 -r_{23}}\qquad r_{13}=\frac{4\, \kappa^2\, r_{23}}{3\,\kappa^2 -r_{23}} \nn\\
Q_0&=& 3 \qquad Q_1=Q_2=\kappa^2\qquad Q_3=2\,\kappa^2  \nn\\
 r_{23} &<& \frac{9-\sqrt{57}}{2}\kappa^2.
\eea
 
 }
 
 \item{  $\kappa_1=\kappa_2=\kappa$, $\kappa_3=2\,\kappa$, $g=(4\,\kappa)^{-1}$
  
\bea
 k^I {}_i&=&
\left(
\begin{array}{ccc}
0 & -\kappa   & 2\,\kappa  \\
 \kappa & 0  & -\kappa \\
 -2\,\kappa & 2\,\kappa &  0
\end{array}
\right)  \qquad r_{12}=\frac{8\, \kappa^2\, r_{23}}{12\,\kappa^2+r_{23}}\qquad    r_{13}=\frac{8\, \kappa^2\, r_{23}}{6\,\kappa^2-r_{23}}.\nn\\
Q_0&=& 3 \qquad Q_1=Q_2= Q_3=2\,\kappa^2\nn\\
r_{23}&<&\kappa^2\left(-11+\sqrt{145}\right).
 \eea
 
}

 \end{itemize}
We notice that in the two examples considered here distances are bounded by triangle inequalities leading to moduli spaces of finite volume.

 \section{ Conclusions}

 We have constructed explicit examples of micro{-}state geometries of four-dimensional black holes.  Following Bena, Gibbons and Warner, we have considered solutions consisting of half-BPS D-brane atoms  with centers in $\mathbb{R}^3$.  Charges  and positions  of the centers are constrained  by the bubble equations that ensure that the metric uplifts to a horizon-less and CTC free metric in five dimensions and  boundary conditions that grant the match of the fuzzball and by the black hole geometries at infinity.  As a result, divergences coming from curvature singularities in the four dimensional metric  are compensated by the singular behaviours of the scalars and gauge fields, leading to a finite (higher-derivative) string effective action. 
 
 We have considered the case of three centers in some details and found that there are two broad classes of solutions. Scaling solutions and non-scaling ones. 
 The moduli space of scaling solutions  is described by a disjoint union of 12 one-dimensional components spanned by a single parameter (up to rigid rotations and translations of the systems) describing the distances between the centers. These solutions carry zero angular momentum and admit a microscopic description in terms of intersecting D3-branes along the lines of \cite{Bianchi:2016bgx}. Non-scaling solution are  described by disjoint unions of one-dimensional components with (generically) a finite volume bounded by the charges of the system.

 \section*{Acknowledgements}
 
 We would like to thank  A.~Addazi, I.~Bena, M.~Bertolini, D. ~Consoli, G.~Dall'Agata, S.~Giusto, A,~Iannuzzi, O.~Lunin, S.~Mathur, R.~Russo, S.~Trapani, M.~Trigiante and G.~Veneziano for very useful discussions and comments on the manuscript. We would like to thank the MIUR-PRIN contract 2015MP2CX4002 ``Non-perturbative aspects of gauge theories and strings'' and the  University of Rome Tor Vergata `Uncovering Excellence' grant ``StaI - String Theory and Inflation'' for partial support. L.~P. would like to thank Queen Mary University of London for hospitality during completion of this work.
 
 \appendix
 
\section{ The ten dimensional solution and its 4d reduction  }
\label{appN1}
In this appendix we collect some details on the dimensional reduction down to four dimensions of the eight harmonic family of BPS solutions describing a general system of intersecting D3-branes on $T^6$.

\subsection{The ten dimensional solution }
 
The eight harmonic family of BPS solutions associated to D3-branes intersecting  on $T^6$ is characterised by  a metric  $g_{MN}$ and a four form Ramond field $C_4$ of the form \cite{Bianchi:2016bgx}
\begin{align}   \label{ansatz}
ds^2  &= g_{\mu\nu} \,dx^\mu\, dx^\nu+\sum_{I=1}^3 h_{mn}^I \,dy_I^m\, dy_I^n\nn \\
C_4&=  C_{\m, mnp}\,dx^\m\wedge dy_1^m\wedge dy_2^n\wedge dy_3^p,
\end{align}
with $\mu=0,\ldots 3$, $m=1,2$.   $x^\mu$ are the coordinates along the four-dimensional space time and $y^m_I=(y_I,\tilde y_I)$ span a $T^2\times T^2\times T^2$ torus with $I=1,2,3$ labelling the three two-torus. More precisely we write \footnote{
  In matrix form  
\eq
h^{I}_{mn}=  \frac{1}{\Imm U_I}\pmat 1 & \text{Re}\, U_I\\ \text{Re}\, U_I& | U_I|^2\fpmat,    \quad\quad 
h^{mn}_I=  \frac{1}{\Imm U_I} \pmat | U_I|^2 &-\text{Re}\, U_I\\-\text{Re}\, U_I & 1\fpmat.
\feq
  }
  \begin{align}
ds^2  &=-e^{2U}\pt{dt+w}^2+e^{-2U}\sum_{i=1}^{3}dx^2_i+\sum_{I=1}^3     {1\over  \text{Im}\, U_I  } \left| dy_I+U_I d\tilde y_I \right|^2 \nonumber\\
C_4&=  A_{\Lambda } \gamma^\Lambda    =A_a \gamma^a + A_{\dot a}\,  \gamma^{\dot a}
\end{align}
with  $\alpha^\Lambda$ one forms in four dimensions and $\gamma_\Lambda$ three forms in the internal torus. $\Lambda=(mnp)=(a,\dot a)$ is a collective index labelling the 8 different 
three cycles $[mnp]$ on $T^6$ entering in the solution
 \begin{align} 
 \gamma^a &=(dy_1 \wedge dy_2\wedge dy_3,  d \tilde y_1 \wedge dy_2\wedge dy_3,     dy_1 \wedge d\tilde y_2\wedge dy_3,dy_1 \wedge dy_2\wedge d\tilde  y_3         )  \nn\\
\gamma^{\dot a} &=( d\tilde y_1 \wedge d\tilde y_2\wedge d\tilde y_3,  d  y_1 \wedge d\tilde y_2\wedge d\tilde y_3,     d\tilde y_1 \wedge d y_2\wedge d\tilde y_3,d\tilde y_1 \wedge d\tilde y_2\wedge d y_3         ) \end{align}
 All functions entering in the metric and four form can be written in   terms 
  of eight harmonic functions 
    \eq
  \{ V, L_I, K_I, M \} \label{harmh}
   \feq   
  on $\mathbb{R}^3$ or equivalently in terms of the following combination  
\begin{align}
Z_I&=L_I+\frac{|\ve_{IJK}|}{2}\frac{K_JK_K}{V},\nn\\
\m&=\frac{M}{2}+\frac{L_IK_I}{2V}+\frac{|\ve_{IJK}|}{6}\frac{K_IK_JK_K}{V^2}. 
\end{align}
 with $\epsilon_{IJK}$ characterising the triple  intersections  between two cycles on $T^6$.
 One finds
  \begin{align}  
 e^{-4U}& = Z_1Z_2Z_3V-\m^2V^2, \nn\\
 U_I & =\text{Re}\, U_I+i\Imm U_I = -b_I+i\pt{Ve^{2U}Z_I}^{-1}   \quad\quad   b_I={K_I\over V} -\frac{\m}{Z_I},\nn\\
A_a & =( \al, \al^I-b^I\al )\nn \\
 A_{\dot a} & 
 =\left(\beta-b^1b^2b^3\al-\beta^I b_I+\half|\ve_{IJK}|\al^I b^J b^K   , ~ \beta^I+   |\ve_{IJK}|\pt{\al b^Jb^K-2b^J\al^K}    \right)   
\end{align}
Finally the one-forms $\alpha,\alpha^I,\beta,\beta^I$ are defined in terms of the eight harmonic function via
\begin{align}
\al&=w_0-\m V^2 e^{4U}\pt{dt+w},\nn\\
\al_I&=-\frac{dt+w}{Z_I}+b_I w_0+w_I,\nn\\
\beta &=-v_0+\frac{e^{-4U}}{V^2 Z_1Z_2Z_3}\pt{dt+w}-b_Iv_I+b_1b_2b_3 w_0+\frac{|\ve_{IJK}|}{2}b_Ib_J w_K,\nn\\
\beta_I&=-v_I+\frac{|\ve_{IJK}|}{2}\pg{\frac{\m \pt{dt+w}}{Z_JZ_K}+b_Jb_K w_0+2b_Jw_K} 
\end{align}
and  
 \begin{align} \label{eq: campi hodge harm}
*_3d w_0&=dV, \quad *_3dw_I=-dK_I , \quad *_3dv_0=dM, \quad 
*_3dv_I=dL_I \nn\\
*_3dw&=\half\pt{VdM-MdV+K_IdL_I-L_IdK_I}
%Vd\m-\m dV-VZ_IdP_I,&e^{-4U} 
\end{align}

\subsection{The four dimensional model}

 After reduction to four dimensions, the ten dimensional solution can be viewed as a solution of a ${\cal N}=2$ truncation of maximal supersymmetric
   supergravity involving the gravity multiplet and three vector multiplets. The four dimensional model arises as a dimensional reduction of the ten dimensional lagrangian
   \eq
\lagr  = \sqrt{g_{10} }  \left( R_{10}  -\frac{1}{4 {\cdot} 5!}F_{M_1\ldots M_5}F^{M_1\ldots M_5} \right)  \label{l10}
\feq 
   Plugging  the ansatz (\ref{ansatz})  into ( \ref{l10}) and taking all fields varying only along the four-dimensional spacetime one finds
  \eq
\lagr = \sqrt{g_4} \left(R_4-\sum_{I=1}^3 \frac{\partial_\m  U_I\partial^\m \bar{U}_I}{2\pt{\Imm U_{I}}^2}-\frac{1}{4{\cdot} 2!}F_{\m\n,\Lambda}F^{\m\n,\Lambda}\right)
\feq
 It is convenient to introduce a metric $\mathcal{H}^{\Lambda\Sigma}$ and its inverse to raise and lower the $\Lambda$ indices. One writes
\eq
\mathcal{H}^{abc,def}=h_1^{ad}\, h_2^{be}\, h_3^{cf}
\feq
or in matrix form
\eq
\mathcal{H}^{\La\Si}=\pmat \mathcal{H}_1& \mathcal{H}_2\\ \mathcal{H}_2^T & \mathcal{H}_3 \fpmat,
\feq
 with $\mathcal{H}_1^{ab}$,  $\mathcal{H}_2^{a\dot a}$, $\mathcal{H}_3^{\dot a \dot b}$   $4\times 4$ matrices. 
The self-duality condition of the five form field in ten dimensions  
 \eq
F_{\m\n abc}=\frac{\sqrt{g_{10}}}{2}  \ve_{\m\n\rho\si abc def}F^{\rho\si def},
\feq
reduces to 
\eq
F_{a bc}=\ve_{abc def}  \widetilde{F}^{def}   \quad \Leftrightarrow    \quad  F_\Lambda=\ve_{\Lambda \Sigma} \widetilde{F}^{\Sigma} 
\feq
with  $\widetilde{F}=* _4 F$ and   $\ve_{\Lambda \Sigma}$ an block off-diagonal antisymmetric matrix with the only non-trivial components
 \eq
\ve_{0\dot{0}}=-\ve_{I\dot{I}}=-\ve_{\dot{0} 0}=\ve_{\dot{I}I}=1  
\feq
In components
\eq\label{selfd}
F_{a}=\ve_{a \dot{a} }\widetilde{F}^{\dot a}   \quad\quad      F_{\dot a}=\ve_{\dot{a}  a }\widetilde{F}^{ a} 
\feq
These self-duality relations can be used to express the components $F_{\dot a}$ in terms of the Poincare' duals of $F_a$. Indeed, inverting the first equation in (\ref{selfd}) one finds\footnote{Here we use $*_4^2=-1$.
}
 \eq\label{selfd2}
       F_{\dot{a}}=-(\mathcal{H}_3^{-1})_{\dot{a}\dot{b}}\left( \ve^{\dot{b}c} \widetilde{F}_{c}+  \mathcal{H}_2^{\dot{b}c}F_c \right) 
\feq
 with $\ve^{\dot{a} a }= {\rm diag} (1,-1,-1,-1)$ the inverse of $\ve_{a \dot{a}  }$. Using these relations one  can write
 \begin{align} 
F_\Lambda F^\Lambda  & ={\cal L}_{stu}+{\cal L}_{top}
 \end{align}
 with $ {\cal L}_{top} =  -2\epsilon^{a\dot a}  \widetilde{F}_{\dot a} F_a$ a total derivative, 
\begin{align}
{\cal L}_{stu} & = 2 ( F_a\mathcal{I}^{ab}F_b+ F_a\mathcal{R}^{ab} \widetilde{F}_b)
\end{align}
and \footnote{Equivalently
\begin{align}
\mathcal{I}^{ab}&\equiv\mathcal{H}_1^{ab}+\ve^{a\dot{b}}(\mathcal{H}_3^{-1})_{\dot{b}\dot{c}}\ve^{\dot{c}b}-\mathcal{H}_2^{a\dot{b}}(\mathcal{H}_3^{-1})_{\dot{c}\dot{d}}\mathcal{H}_2^{\dot{d}b} \nn\\
\mathcal{R}^{ab}&\equiv \ve^{a\dot{b}} (\mathcal{H}_3^{-1})_{\dot{b}\dot{c}}\mathcal{H}_2^{\dot{c}b}+\mathcal{H}_2^{a\dot{b}}(\mathcal{H}_3^{-1})_{\dot{b}\dot{c}}\ve^{\dot{c}b} \nn 
\end{align}
}
\begin{align*}
\mathcal{I}^{ab}&= stu \pmat 1+\frac{\si^2}{s^2}+\frac{\tau^2}{t^2}+\frac{\nu^2}{u^2} & -\frac{\si}{s^2}&-\frac{\tau}{t^2}&-\frac{\nu}{u^2} \\
-\frac{\si}{s^2}  & \frac{1}{s^2} &0 &0 \\
-\frac{\tau}{t^2}  & 0 &\frac{1}{t^2}&0\\
-\frac{\nu}{u^2}  & 0 & 0 &\frac{1}{u^2}\fpmat		\quad
\mathcal{R}^{ab}=\pmat 2\si\tau\nu & -\tau\nu&-\sigma\nu& -\sigma\tau \\
-\tau\nu & 0 & \nu &\tau \\
-\sigma\nu  &\nu &0&\sigma\\
-\sigma\tau   &\tau &\sigma &0 \fpmat \num
\end{align*}
where
\begin{align}
U_I&=(\si+is,\tau+it,\nu+i u ) 
\end{align}
 Discarding the total derivative term, the four-dimensional  Lagrangian can then be written as
\eq
\lagr =  \sqrt{g_4}\left( R_4-\sum_{I=1}^3 \frac{\partial_\m U_I\partial^\m \bar{U}_I}{2\pt{\Imm U_{I}}^2}-\frac{1}{4}F_a \mathcal{I}^{ab}F_{b}-\frac{1}{4}F_{a}\mathcal{R}^{ab} \widetilde{F}_{b} \right)
\feq
The equations of motion read
\begin{align}
 & R_{\m\n} -\ft12 g_{\m\n} =   {1\over 2\pt{\Imm U_{I}}^2 } \left( \partial_\m U_I \partial_\n U_I - \half g_{\m\n} (\partial U_I) ^2 \right) 
 + \ft12 {\cal I}^{ab} \left( F_{a\mu \sigma}  F_{b\n}^\sigma-\quar  g_{\m\n}F_a\, F_b \right) \nn \\
& \quad\quad\quad\quad   + \ft12 {\cal R}^{ab} \left( F_{a\mu \sigma}  \tilde F_{b\n}^\sigma-\quar  g_{\m\n}F_a\, \tilde F_b \right) \nn\\
 & \nabla_\m\pg{    {\cal I  }^{ab} F_b^{\m\n}+ {\cal R}^{ab} \tilde F_b^{\m\n} }=0  \nn\\
&  -\nabla_\m \frac{\nabla^\m U_I}{\pt{\Imm U_I}^2}= {\rm i}   \frac{\partial_\m U_I\partial^\m \bar{U}_I}{\pt{\Imm U_{I}}^3}+\frac{1}{2}F_a  {\partial \mathcal{I}^{ab} \over \partial \bar{U}_I} F_{b}+\frac{1}{2}F_{a}{\partial \mathcal{R}^{ab} \over \partial \bar{U}_I}   \widetilde{F}_{b}   \label{eom}
\end{align}

\subsection{The basic solutions}
\label{appN1basic}

A family of supersymmetric solutions to equations (\ref{eom}) is given in \cite{Bianchi:2016bgx}.  These solutions can be viewed as made of three different types of solutions, referred as K, L or M. In the following  we display a representative of solution in each type.

\subsubsection{L solutions}
 
 The L class of solutions can be represented by the choice 
\begin{align}
V &\equiv L\pt{x},\quad M=K_I=0,\quad L_I=1 \quad \Rightarrow \quad Z_I=1,\quad \m=0\nn\\
 \mathcal{I} &=\diag{L^{-3/2},L^{-1/2},L^{-1/2},L^{-1/2}}		\qquad
\mathcal{R}=0
\end{align}
%that leads to the effective Lagrangian 
% \eq
%\lagr_L= \sqrt{g} \left(R_4-\frac{3}{8}\frac{\pt{\partial L}^2}{L^2}-\ft{1}{4} \,L^{-3/2}\, F^2_0 \right)
%\feq 
% and equations of motion
%\begin{align}
%& R_{\m\n} -\ft12 g_{\m\n} =\frac{3L^{-2}}{8}\pt{\partial_\m L\partial_\n L-\half  g_{\m\n} \pt{\partial L}^2}
%+\frac{L^{-3/2}}{2}  \left( F_{0\mu \sigma}  F_{0\n}^\sigma-\quar  g_{\m\n}F_0^2 \right) \nn\\
%&  \nabla_\m \pt{L^{-3/2}F_0^{\m\n}}=0\nn\\
%&  \nabla_\m\pt{\frac{\nabla^\m L}{L^2}} =-\half L^{-5/2}F_0^2-\frac{\nabla_\m L\nabla^\m L}{L^3}.
%\end{align}
  The solution can be written as
\begin{align}
ds^2 &=-L^{-{1\over 2}}dt^2+L^{1\over 2} \sum_{i=1}^3 dx^2_i \nn\\
 A_0  &=  w_0   \qquad   *_3 d w_0=dL \nn\\
 U_1&=U_2=U_3={\rm i}\, L^{-{1\over 2}}
\end{align}

\subsubsection{K solutions}
 
 The K solutions correspond to the choice
\begin{align}
K_3 &=-M\equiv K\pt{x},\quad L_I=V=1,\quad K_1=K_2=0  \quad \Rightarrow \quad Z_I=1,\quad \m=0 \nn\\
\mathcal{I} &=\pmat 1+K^2 & 0&0&K \\
0 & 1 &0 &0 \\
0 & 0 &1&0\\
K & 0 & 0 &1\fpmat		\qquad ~~~~~~~~~~
\mathcal{R} =\pmat 0 &0&0&0 \\
0 & 0 & -K &0 \\
0 & -K &0&0\\
0 &0 & 0 &0 \fpmat \num
\end{align}
%leading to the effective Lagrangian 
%\eq
%\lagr_K= \sqrt{g} \left(  R_4-\partial_\mu K^2-\ft14 \sum_{I=1}^2 \left(  F_I^2- K\,  F_I\Ft_I \right) \right) 
%\feq
%and equations of motion
%\begin{align}
% & R_{\m\n} -\ft12 g_{\m\n} =   \half\left( \partial_\m K\partial_\n K- \half g_{\m\n} (\partial K) ^2 \right) + \ft12 \sum_{I=1}^2 \left( F_{I\mu \sigma}  F_{I\n}^\sigma-\quar  g_{\m\n}F_I^2 \right) \\
% & \nabla_\m\pg{\pt{1+K^2}F^{\m\n}_I-K\Ft^{\m\n}_I}=0 \quad\quad I=1,2\nn\\
%&  \nabla_\m \nabla^\m K=-\half F_I \widetilde{F}_I=0,
%\end{align}
 
The solution is given by
\begin{align}
ds^2&=- \pt{dt+w}^2+\sum_{i=1}^3 dx^2_i,  \nn\\
U_1 &=U_2={\rm i} \quad\quad   U_3=-K+{\rm i} \nn\\
A_0&=A_3=0   \quad\quad\quad \quad  A_1=A_2= -w       \quad\quad    *_3 dw =-dK 
\end{align}

\subsubsection{M solutions}

The M solutions correspond to the choice
\begin{align}
K_2 &=M\equiv M\pt{x},\qquad L_I=V=1,\qquad K_1=K_3=0  \quad \Rightarrow \quad
\mu=M \quad Z_I=1,\quad    \nn\\
\mathcal{I} &=a^{3/2}\pmat 1+2\frac{M^2}{a} & -\frac{M}{a}&0&-\frac{M}{a} \\
-\frac{M}{a} & a^{-1} &0 &0 \\
0 & 0 & a^{-1}&0\\
-\frac{M}{a} & 0 & 0  & a^{-1}\fpmat		\quad\quad
\mathcal{R}=\pmat 0 & 0&-M^2& 0 \\
0 & 0 & M&0 \\
-M^2  & M &0&M\\
0 & 0 & M  &0 \fpmat \num
\end{align}
with $a=1-M^2$.
The solution is given by
\begin{align}
ds^2&=- { dt^2 \over \sqrt{1-M^2} }+\sqrt{1-M^2} \sum_{i=1}^3 dx^2_i,  \nn\\
U_1 &=U_3=M+{\rm i} \, \sqrt{1-M^2}  \quad\quad U_2={\rm i} \, \sqrt{1-M^2}  \nn\\
A_0&=-\frac{M\, dt}{1-M^2}  \quad\quad  A_1=A_3= -\frac{dt}{1-M^2}   \quad\quad  A_2=w_2 \qquad   *_3 dw_2=-d M
\end{align}

\subsection{Sub-family of solutions}
\label{appN2}

For completeness, we list some interesting sub-families of solutions included in the eight-harmonic class. 

 \subsubsection{No scalars: IWP solution }
 
 Einstein-Maxwell theory can be embedded in four dimensional supergravity by restricting to geometries with a trivial internal square metric 
  \bea
V e^{2U} Z =1 \qquad  b_I=0 
\label{nointernal}
\eea
  These equations can be solved in terms of two harmonic functions $\text{Re}\, H$ and $ \text{Im}\, H$ via the identifications
\bea
V &=& L_1=L_2=L_3=\text{Im}\,H  \nn\\
-M&=& K_1  = K_2=K_3=\text{Re}\, H
\eea
The general solution reduces to\footnote{ In our conventions the Einstein-Maxwell lagrangian reads
\be 
{\cal L}= \sqrt{g} \left[   R-\ft{1}{4 } \, F^2 \right]
\ee
}
\bea
ds^2 &=& -  \left| H  \right|^{-2} \, (dt+w)^2 + \left| H  \right|^2 \,  d \vec x^2 \nn\\
A_0 &=&   w_0 -{ \text{Re}\,H \over \left| H  \right|^2  } \, (dt +w)  \nn\\
    A_1 &=& A_2=A_3=  w_1 -{ \text{Im}\,H \over \left| H  \right|^2  } \, (dt +w)\label{iwp}  
\eea
with $H$ a complex harmonic function
 \be
 H=\text{Re}\, H+\ii\,\text{Im}\, H \qquad\quad \quad  \nabla^2 H= 0
 \ee
  and $w$ and $w_0$, $w_1$ one forms defined as
   \bea
d w&=&  {\rm i} *_3  [H  dH^*-H^* dH ] \nn\\
d w_0 &=& *_3  d \, \text{Im}\, H   \qquad    d w_1 = *_3  d \, \text{Re}\, H \label{wb}
\eea
 We notice that the contribution to the stress energy tensor of gauge fields $A_I$ exactly match that of $A_0$ , so we can replace the four gauge fields by a 
 single one given by
\be
 A =  2 w_0 -{ 2\text{Re}\,H \over \left| H  \right|^2  } \, (dt +w)  \nn\\
\ee
   The resulting solution is known in the General Relativity literature as IWP ( after Israel, Wilson and Perjes  \cite{Perjes:1971gv,Israel:1972vx} ) and
  includes very well known examples of solutions of Maxwell-Einstein gravity:
\begin{itemize}
\item{$AdS_2\times S^2$. The harmonic function $H$ reads\footnote{ Global coordinates are defined by
\be
(x_1,x_2,x_3)  = \left(  \sqrt{(\rho^2 +L^2) (1-\chi^2) }  \, \cos\phi, \sqrt{(\rho^2 +L^2)(1-\chi^2) }\, \sin\phi,  \rho\,  \chi \right)  
\ee
with $\rho \in (-\infty,\infty)$,  $\chi \in [-1, 1]$, $ \phi \in [0,2 \pi]$. These coordinates cover twice the flat space  with
 the points $(\rho,\chi)$ and $(-\rho,-\chi)$  identified.
}
\be
H={1\over \sqrt{ x_1^2+x_2^2+(x_3-{\rm i} L)^2 } }
\ee
The geometry is regular everywhere. An infinite class of regular IWP geometries, obtained as bubbling of $AdS_2\times S^2$ and parametrised by a string profile function has been recently constructed in \cite{Lunin:2015hma}. 
}

\item{ Kerr-Newman solution with  $M=Q=q$, $P=0$, $J=q L$. The harmonic function $H$ reads 
\be
H=1+{q\over \sqrt{ x_1^2+x_2^2+(x_3-{\rm i} L)^2 } }
\ee
The geometry has a naked curvature singularity at the zero of $H$.   
} 

\item{ Reissner-Nordstrom with  $M=Q=q$, $P=J=0$. The harmonic function $H$ reads 
\be
H=1+{q \over \sqrt{ x_1^2+x_2^2+x_3^2 } }
\ee
The geometry has  a curvature singularities at the zero of $H$.
   
}

\item{ Charged Taub-NUT with  $M=Q=b_1$, $P=-b_2$, $J=0$. The harmonic function $H$ reads 
\be
H=1+{b_1+{\rm i} \, b_2\over \sqrt{ x_1^2+x_2^2+x_3^2 } }
\ee
The geometry has  no curvature singularities but it has a Dirac-Misner string-like singularity.
   
}

\end{itemize}

\subsubsection{One complex scalar:  SWIP solutions}

Next, we consider a solutions with single active scalar field, let us say $U_1$,  with $U_2=U_3={\rm i}$. These conditions can be solved in terms of two 
complex harmonic functions $H_1$, $H_2$ after the identification
\begin{align}
%K_1 &=-M\equiv -\text{Re}\,H_1,\qquad L_2=L_3\equiv \Imm H_1,\qquad L_1=V\equiv \text{Re}\,H_2,\qquad K_2=K_3\equiv \Imm H_2\label{klmswip}
L_1 &=V = \text{Re}\,H_2,\qquad L_2=L_3\equiv \Imm H_1\qquad K_1=-M = \text{Re}\, H_1\,\qquad K_2=K_3\equiv -\Imm H_2\label{klmswip}
\end{align}
%leading to a four dimensional model with gauge kinetic functions
%\begin{align}
%\mathcal{I} &=\pmat \frac{\ps{H_1}^2}{\Imm \pt{H_1\bar{H}_2}} & -\frac{\text{Re}\, \pt{H_1\bar{H}_2}}{\Imm \pt{H_1\bar{H}_2}}&0&0 \\
% -\frac{\text{Re}\,\pt{H_1\bar{H}_2}}{\Imm \pt{H_1\bar{H}_2}} & \frac{\ps{H_2}^2}{\Imm \pt{H_1\bar{H}_2}} &0 &0 \\
%0 & 0 & \frac{\Imm \pt{H_1\bar{H}_2}}{\ps{H_2}^2}&0\\
%0 & 0 & 0  & \frac{\Imm \pt{H_1\bar{H}_2}}{\ps{H_2}^2}\fpmat \qquad 
%\mathcal{R}&=\pmat 0 & 0&0& 0 \\
%0 & 0 & 0&0 \\
%0 & 0 &0&\frac{\text{Re}\, \pt{H_1\bar{H}_2}}{\ps{H_2}^2}\\
%0 & 0 & \frac{\text{Re}\, \pt{H_1\bar{H}_2}}{\ps{H_2}^2}  &0 \fpmat .
%\end{align}
%Pluggin (\ref{klmswip}) into (\ref{pam}) one finds
%\begin{align}
%e^{-2U}&=\Imm\pt{H_1\bar{H_2}},\qquad \mu=\frac{\Imm \pt{H_1 \bar{H_2}}}{ (\text{Re}\,H_2)^2 }  \nn\\
%P_1&=-\frac{\text{Re}\,H_1}{\text{Re}\,H_2}, \qquad P_2=P_3=\frac{\Imm H_2}{\text{Re}\,H_2},\qquad Z_1=\frac{\ps{H_2}^2}{\text{Re}\,H_2},\qquad Z_2=Z_3=\frac{\Imm \pt{H_1 \bar{H_2}}}{\text{Re}\,H_2} 
%\end{align}
 For this choice the general solution reduces to
\begin{align}
ds^2&=-  \pq{\Imm\pt{H_1\bar{H}_2}}^{-1}\pt{dt+w}^2 +\Imm\pt{H_1\bar{H}_2} \sum_{i=1}^3 dx^2_i, \nn \\
 U_1 &=\frac{H_1}{H_2},\qquad U_2=U_3={\rm i},  \nn\\
A_0&=w_0+\frac{\Imm H_2}{\Imm\pt{H_1\bar{H_2}}}\pt{dt+w},  \qquad  A_1=w_1-\frac{\Imm H_1}{\Imm\pt{H_1\bar{H_2}}}\pt{dt+w}  \nn \\
  A_2&=A_3=w_2-\frac{\text{Re}\, H_2}{\Imm\pt{H_1\bar{H_2}}}\pt{dt+w}\nn\\
  *_3 dw &=-\,\text{Re}\, \pt{H_1 d\bar{H_2}-\bar{H}_2 dH_1}\nn\\
*_3d w_0&=\text{Re}\, dH_2,\qquad *_3dw_1=-\text{Re}\, dH_1,\qquad *_3dw_2=\Imm dH_2.
\end{align}
  The IWP  class  corresponds to the choice $H_1={\rm i} H_2= H$. See for instance \cite{Ortin:2015hya} for more information on the SWIP solution.
 
  \subsubsection{Two complex scalars}
This solution corresponds to the choice
\begin{align}
  L_3=L_2    \qquad    K_3=K_2  \label{klmswip2}
 \end{align}
 leading to 
 \bea
 Z_1 &=&   L_1+{K_2^2\over V}  \qquad\qquad    Z_2 =Z_3=   L_2+{K_1 K_2\over V} \nn\\
 \mu &=&  { M \over 2} + { K_1 \,K_2^2   \over   V^2} + { K_1\, L_1 \over   2 V}+
 {  K_2 L_2 \over V}
 \eea
The solution reads
\begin{align}
ds^2&=-e^{-2U}\pt{ dt+w}^2+e^{2U}\sum_{i=1}^3 dx^2_i,   \qquad e^{-4 U} =Z_1 Z_2^2 V-\mu^2 V^2\nn\\
U_1 &=-b_1+{\rm i} (e^{2U} V Z_1)^{-1}   \qquad   U_2=U_3=-b_2+{\rm i} (e^{2U} V Z_2)^{-1} \nn\\
A_0&=w_0-\m V^2 e^{4U}\pt{dt+w}     \qquad      A_1 =w_1   + V e^{4U} \, \pt{dt+w}   
  \left(  Z_2^2  -K_1\, \m     \right) \nn\\
  A_2 & =A_3 =w_2  + V e^{4U} \, \pt{dt+w}   
  \left(  Z_1 Z_2  -K_I \, \m     \right)
\end{align}
with
\begin{align}
 *_3 dw &= \ft12(V d M-M dV+K_1 dL_1 -L_1 dK_1+ 2 K_2 dL_2-2L_2 dK_2   )      \nn\\
 *_3d w_0&=dV,\qquad *_3dw_1=-dK_1 \qquad *_3dw_2= -dK_2  
\end{align}
and
\begin{align}
b_1 &={ K_1 L_1-2 K_2 L_2-M V \over 2(K_2^2+V L_1) }   \qquad
b_2 =b_3 =-{ K_1 L_1 +M V \over 2(K_1 K_2+V L_2 )}  
\end{align}
The SWIP solution is recovered for $L_1=V$ and $K_1-M$ after the identifications (\ref{klmswip}).

%\bibliography{references_D3scalar}
%\bibliographystyle{abe}

% $$$$$$$$$$$$$$$$$$$$$$$$$$$$$$$$$$$$$

\providecommand{\href}[2]{#2}\begingroup\raggedright\endgroup

% $$$$$$$$$$$$$$$$$$$$$$$$$$$$$$$$$$$$$

\end{document}